\documentclass[12pt,preprint]{aastex}

\shorttitle{Predicted and empirical radii of RR Lyrae stars}
\shortauthors{Marconi et al.}

\begin{document}

\title{Predicted and empirical radii of RR Lyrae stars}

\author{Marconi, M.\altaffilmark{1}, 
Nordgren, T.\altaffilmark{2},
Bono, G.\altaffilmark{3}, 
Schnider, G.\altaffilmark{2}, and 
Caputo, F.\altaffilmark{3}} 

\altaffiltext{1}{INAF - Osservatorio Astronomico di Capodimonte, via 
Moiarello 16, 80131 Napoli, Italy; marcella@na.astro.it}
\altaffiltext{2}{Department of Physics, University of Redlands, 1200 East 
Colton Avenue, Redlands, CA 92373, USA; Tyler\_Nordgren@redlands.edu, 
her\_own\_wings@hotmail.com} 
\altaffiltext{3}{INAF - Osservatorio Astronomico di Roma, Via Frascati 33, 
00040 Monte Porzio Catone Italy; bono,caputo@mporzio.astro.it}

\date{\centering drafted \today\ / Received / Accepted }

\begin{abstract}
We present new theoretical Period-Radius-Metallicity (PRZ) relations for 
RR Lyrae stars. Current predictions are based on a large set of nonlinear,
convective models that cover a broad range of chemical abundances and 
input parameters. We also provide new and homogeneous estimates of 
angular diameters for a sample of field RR Lyrae stars using a recent 
calibration of the Barnes-Evans surface brightness relation. Predicted 
and empirical radii are, within the errors, in reasonable agreement, 
but in the short-period range the latter present a larger scatter. 
As a working hypothesis we suggest that this discrepancy might be 
due to the occurrence either of nonlinear features such as bumps or 
a steep rising branch. New distance determination for RR Lyr itself is 
in very good agreement with HST trigonometric parallax and with 
pulsation parallax.  
\end{abstract}


\maketitle 

\section{Introduction}

RR Lyrae stars are widely adopted not only as tracers of old, low-mass 
stellar populations but also as standard candles to estimate Galactic 
and extragalactic distances. They are ubiquitous across the Galaxy and 
they have been detected in all stellar systems that host a well-defined 
old population. This means that they can be adopted to constrain the 
intrinsic accuracy of current primary distance indicators, such as 
classical Cepheids, Tip of the Red Giant Branch, and Main Sequence 
fitting. These are the reasons why a countless number of theoretical 
and empirical investigations have been devoted to the RR Lyrae distance 
scale (Caputo et al. 1999; Bono et al. 2002,2003; Cacciari \& 
Clementini 2003; Walker 2003; Catelan, Pritzl, \& Smith 2004; 
Gratton et al. 2004).  

Even though a paramount observational effort has been devoted to obtain 
Baade-Wesselink (BW) distances to a sizable sample of cluster and field 
RR Lyrae (Cacciari et al. 1989; Clementini et al. 1990; 
Carney et al. 1992, hereafter C92; Jones et al. 1992; Storm et al. 1994a,b) 
we still lack detailed empirical and theoretical Period-Radius 
(PR) relations for RR Lyrae stars. Note that radii are a by product 
of the BW method. There is only one exception, Burki \& Meylan (1986) 
derived, using BW measurements, an empirical PR relation for 
Type II Cepheids that according to the authors could also be 
applied to RR Lyrae stars. 

The same outcome applies to theoretical models. Pulsation properties 
of RR Lyrae stars have been investigated using both linear 
and nonlinear models, we still lack detailed theoretical predictions.
To fill this gap we present new theoretical  
PR relations for RR Lyrae stars based on a detailed and homogeneous set 
of nonlinear pulsation models that cover a wide range of stellar masses 
and chemical compositions. 
We also investigate the dependence of the PR relation on metallicity and 
compare theoretical predictions with empirical radius estimates. 

\section{Predicted and Empirical radii of RR Lyrae}

During the last few years we have been developing an homogeneous theoretical 
scenario for RR Lyrae stars by constructing an extensive grid of nonlinear,
convective models (Bono et al. 1997, 2001, 2003). These models cover a wide 
range of stellar masses (0.53 $\le$ M/M$_\odot$ $\le$ 0.80), 
and chemical compositions (0.24 $\le$ Y(He abundance) $\le$ 0.28; 
0.0001 $\le$ Z (metal abundance) $\le$ 0.02).
The physical and numerical assumptions adopted in the model computations 
have already been discussed in previous papers 
(Bono \& Stellingwerf 1994; Bono et al. 1997; Bono, Castellani, \& Marconi 2000). 
Note that current nonlinear, convective models predict pulsation 
observables such as the variation along the pulsation cycle of luminosity, 
radius, velocity, and temperature to be compared with actual empirical data. 
We estimated for each model the mean radius as a time 
average over the predicted surface radius curve. 
Figure 1 shows predicted radii for fundamental (F) pulsators, the solid 
line displays the linear regression over the entire set of models:  

$$\log R = 0.90 (\pm0.03) + 0.65 (\pm0.03) \log P \;\;\;\; \;\;\;\;\;\sigma=0.03$$ 

where $R$ is the mean RR Lyrae radius (solar units), $P$ the pulsation period (days),
and $\sigma$ the intrinsic dispersion.   
For comparison Fig. 1 also shows the empirical PR relation (dashed line) for Type II 
Cepheids derived by Burki \& Meylan (1986). The vertical error bar is the 
standard deviation of the predicted PR relation. The agreement between 
the empirical relation and current predictions is, within the intrinsic dispersion, 
quite good. This evidence supports the suggestion by Burki \& Meylan concerning 
the similarity between the PR relation of Type II Cepheids and RR Lyrae.    
However theoretical radii present, at fixed period, a substantial 
spread along the best-fit line, thus strongly suggesting the dependence of the 
PR relation on a {\it second parameter}.


To further improve the intrinsic accuracy of the predicted PR relation we accounted  
for metal abundance. The dependence on this parameter is expected, since both theory 
and observations support the evidence that the mean magnitude of RR Lyrae stars 
depends on the metal content. We performed a linear regression over the entire set 
of F models and we found the following Period-Radius-Metallicity (PRZ) relation: 

$$\log R = 0.774(\pm0.009)\, +\, 0.580(\pm0.007) \log P\, -\, 0.035(\pm0.001) \log Z  \;\;\;\;\; \;\;\;\;\;\sigma=0.008$$

where the symbols have their usual meaning. 
Data plotted in the bottom panel of Fig. 2 show that the inclusion of the metallicity 
term causes a decrease in the intrinsic dispersion from 0.03 to 0.008 dex. To supply 
a homogeneous theoretical scenario for RR Lyrae radii we also estimated the PRZ 
relation for first overtone (FO) pulsators (see top panel of Fig. 2), and we 
found:    

$$\log R = 0.883(\pm0.004)\, +\, 0.621(\pm0.004) \log P\, -\, 0.0302(\pm0.001) \log Z \;\;\;\;\; \;\;\;\;\; \sigma=0.004$$

The reason why we estimated an independent PR relation for FOs is twofold: 
{\em i))} the width in temperature of the FO instability strip is narrower when 
compared with the F one. This means that the FO PR relation presents a smaller 
intrinsic dispersion when compared with the F one.  
{\em ii))}  FO pulsators are systematically hotter than F ones. This means that 
the predicted FO PR relation is marginally affected by uncertainties in the 
treatment of convection.


To validate current predictions we collected a sample of RR Lyrae stars 
for which are available accurate photometric and spectroscopic data, 
namely the BW sample (Bono et al. 2003, hereafter B03). From these, we calculated 
angular diameters using the latest calibration (Nordgren et al. 2002) of the 
Barnes-Evans surface brightness relation (Barnes \& Evans 1976). Together with 
radial velocities from spectroscopic data, we determined linear radii 
(and distances) for selected RR Lyrae (see Table 1).
While all of these stars already had radii determined by various authors 
(Jones et al. 1992, C92, and see references in Table 1) this 
new sample has the benefit of uniformity and the most recent stellar 
interferometric angular diameter measurements. 

We compiled published V and K photometry for the stars in Table 1 (see 
references in column 9).
The mean values of these V and K magnitudes agree with those in B03 to within an
average of 0.04 magnitudes (about 0.4\% given an average V and K magnitude of 10).
We performed dereddening of the photometric data
using values for E(B-V) provided in Table 1, and extinction correction
constants from Cardelli et al. (1989).
We used both linear interpolation and polynomial-fitting in order to
calculate K, and thus (V-K) values at those
pulsation phases with V photometry. For RR Lyr itself, we used V and (V-R) photometry.
We calculated angular diameters as a function of pulsation phase for stars in Table 1
using equations (1), (2), (5) and (6) of Nordgren et al. (2002) we obtain:

\begin{center} 
$\log\theta = 0.5734 - 0.2V + 0.246(V-K)$

$\log\theta = 0.5914 - 0.2V + 0.730(V-R)$
\end{center} 

where $\theta$ is the angular diameter in milliarcseconds (mas).
This version of the Barnes-Evans relation was calibrated using
interferometric angular diameter observations of 57 non-variable giant stars. 
Where available, angular diameters were calculated using  V and (V-K) pairs, 
as opposed to V and (V-R), as the former relation has been shown to yield more 
precise results (Fouqu\'{e} \& Gieren 1997).
Using polynomial fitting (with polynomial orders ranging from 9 to 11) we calculated
radial velocities from spectroscopic measurements at the same pulsation phases as the
calculated angular diameters (see references in Column 9 of Table 1). 
We calculated linear displacements of the stellar surface from:

$$\Delta R = - p \int{(V_r - V^\ast)} d\phi $$

where $V^\ast$ is the radial velocity of the center of mass of the star and
$p$ is the pulsation projection factor. The value of $V^\ast$ was found by 
integrating the $V_r$ curve over the entire phase cycle (phase = 0 - 1), and 
demanding that $\Delta R_{0-1}$ = 0. For all but RS Boo, our value of $V^\ast$ 
agrees within the uncertainties with those published by Beers et al. (2000).  
For the pulsation factor, Fernley (1994) argues for $p = 1.38$ for field 
RR Lyrae stars and so we have used this value.

The linear radius ($R_o$) and distance ($d$) for the star is found from
$R_o + \Delta R = 1000 d (\theta/2)$, where the angular diameter $\theta$ 
is in mas and the linear radius and radius displacement are 
in AU (yielding the distance, $d$, in parsecs). From this equation the 
radii (and distances) in Table 1 were calculated using a least-squares fit
to our calculated angular diameters and linear displacements. 

The above equation for $\Delta R$ shows that uncertainties in $V^\ast$ will be
propagated into the uncertainty in radius and distance. For each individual star 
this is a systematic error, but it is an error that is random from star to star 
within the sample.  The uncertainties in radius and distance in Table 1 are the 
random errors, with the additional systematic error due to $V^\ast$ given in 
parentheses. The total error for each star is the quadrature sum of the two 
errors. The comparison of random and systematic errors in Table 1 supports 
the importance of accurate radial velocity measurements in BW type analyses.

It should also be noted that there is an uncertainty in the value for the 
pulsation projection factor $p$.  As the linear radius displacement scales 
linearly with $p$, so too will the final radius and distance. For instance, if one were 
to use $p = 1.30$ (Jones, Carney \& Latham, 1988a,b) instead of our value of $1.38$,
all radii and distances in Table 1 will be smaller by a factor of 1.38/1.30 = 1.08.
For each star in Table 1, our new empirical radius estimate is in agreement, within 
the errors, to that found by the authors of the original published photometry and 
radial velocities (see references in Table 1). This includes the star with the 
largest estimated radius, SS Leo. Fernley et al. (1990) estimated a radius of 
6.63 $R_\odot$, while Jones et al. (1992) estimate a radius of 7.32 $R_\odot$, 
both of which, without better knowledge of their uncertainties, are in general 
agreement with our estimate of $7.2 \pm 0.4 R_\odot$. 

Figure 2 also shows the comparison between predicted and new empirical 
(open circles) radius estimates. Triangles display the radius estimates 
provided by C92.  Homogeneous radius determinations for a larger sample of 
BW RR Lyrae with accurate radial velocity measurements will be provided in 
a forthcoming investigation. 
Observed radii have been plotted using the homogeneous compilation of metal 
abundances provided by Fernley et al. (1998) and listed in column 3) of 
Table 1. The reader interested in a detailed discussion concerning the 
metallicity measurements and the metallicity scale of RR Lyrae stars is 
referred to  Dall'Ora et al. (2004) and to Gratton et al. (2004). 
Data plotted in Fig. 2 indicate that empirical radius estimates are affected by 
large scatter. However, theory and observations are, within current uncertainties, 
in reasonable agreement for periods longer than 0.42 days. The radius measurements 
by C92 do not include individual error estimates, and therefore, 
it is not clear whether the three objects with $P > 0.63$ days present a real 
discrepancy. However, observed radii show a larger scatter when moving toward 
shorter periods. The reason for this drift is not clear, however four shorter 
period RR Lyrae present a well-defined bump along the decreasing branch and 
a steep rising branch in both light and velocity curves (RS Boo, TW Her, 
Jones et al. 1988b; V445 Oph, Fernley et al. 1990; W Crt, Skillen et al. 1993). 
Moreover, DH Peg is a FO RR Lyrae.

\section{Discussion and final remarks}

Recent improvements in optical and infrared interferometry have allowed for
direct evaluation of the accuracy of radii and distances determined
by BW analyses of pulsating stars, in particular Cepheid variables
(Nordgren et al. 2002; Lane et al. 2002; Kervella et al. 2004). Since 
no RR Lyrae star currently has an angular diameter directly measured 
by interferometry, such a comparison of the accuracy of our RR Lyrae 
radii is not possible. However, one may compare the distances calculated 
using our method as a test of the accuracy of our surface brightness
relation, provided there is a known distance to any of the stars in our sample.
Alone of the stars in Table 1, RR Lyr has a distance known from high precision
trigonometric parallax observations. Benedict et al. (2002, hereinafter B02) 
used the Hubble Space Telescope Fine Guidance Sensor to obtain a parallax of 
$\pi = 3.82 \pm 0.2$ mas yielding a distance of $262 \pm 14$ pc. As a check 
of our surface brightness analysis we compare this relatively model independent 
distance to the distance calculated in Table 1: $270 \pm 35$ pc (this uncertainty 
is the quadrature sum of the listed random and 
systematic uncertainties). Our distance is in excellent agreement with the HST 
distance and with the distance obtained using the K-band Period-Luminosity-Metallicity 
($PLZ_{\rm K}$) relation ($260\pm 5$ pc) obtained by B03. This agreement gives us confidence in 
the accuracy of surface brightness relations and their results. It should be noted that
no horizontal branch stars were included in the calibration of the surface brightness 
relation used here (Nordgren et al. 2002). That RR Lyrae radii determined from this calibration 
agree so well with the theoretically computed radii argues that the surface brightnesses of
horizontal branch stars may be well computed from calibrations based on giant stars.

According to this evidence, we compared current distance determinations with distances 
estimated using the RR Lyrae visual magnitude metallicity relation 
($M_V\; vs\; [Fe/H]$) provided by B02. Data plotted 
in the top panel of Fig. 3, show that the relative difference is within an average of 
10\%. As expected, the discrepancy is significantly larger in the short-period range. 
To constrain the intrinsic accuracy of current distances, the bottom panel 
of Fig. 3 shows the relative difference with the distances based on the $PLZ_{\rm K}$
relation. A glance at the data plotted in this panel confirms the quoted result. 

Current findings suggest that predicted and observed radii of RR Lyrae stars are in 
reasonable agreement. The accuracy of empirical estimates do not allow us to constrain 
the plausibility of nonlinear, convective RR Lyrae models. Needless to say, that this 
analysis shall be extended to the entire sample of cluster and field RR Lyrae stars 
for which are available accurate spectroscopic and photometric (optical, NIR) 
measurements. In the future, more precise trigonometric and pulsation parallaxes 
together with new angular diameter measurements, will certainly improve the 
observational scenario not only for radii and distances but also for the 
pulsation factor $p$. 

It is a pleasure to thank an anonymous referee for his/her positive comments 
and suggestions. This work was partially supported by PRIN~2003 and INAF~2003. 


\clearpage 
\begin{deluxetable}{lcccrrrrl}
\scriptsize 
\tablecolumns{9}
\tablewidth{0pt}
\tablecaption{Selected Baade-Wesselink RR Lyrae}
\tablehead{
\colhead{Star} & \colhead{Period} & \colhead{[Fe/H]} & \colhead{E(B-V)} & \colhead{$V^\ast$\tablenotemark{a}} &
\colhead{Phase} & \colhead{Distance} & \colhead{Radius} & \colhead{Ref.}\\
\colhead{} & \colhead{(days)} & \colhead{} & \colhead{} & \colhead{} &
\colhead{} & \colhead{(pc)} & \colhead{(R$_\odot$)} & \colhead{}
}
\startdata
DH Peg\tablenotemark{b}&0.2555&-1.24 & 0.08  & -71 $\pm$ 1& 0.0-0.8  & 470 $\pm$ 60(50) & 3.8 $\pm$ 0.5(0.5) & e \\
RS Boo	& 0.3773 & -0.36 & 0.02  &  -4 $\pm$ 2 & 0.35-0.8 & 770 $\pm$ 20(120) & 4.1 $\pm$ 0.1(0.7) & g \\
TW Her	& 0.3996 & -0.69 & 0.05  &  -5 $\pm$ 2 & 0.1-0.9  &1150 $\pm$ 35(100) & 4.2 $\pm$ 0.1(0.5) & g \\
V445 Oph& 0.4000 & -0.19 & 0.27  & -18 $\pm$ 1 & 0.0-1.0  & 700 $\pm$ 15(30)  & 4.4 $\pm$ 0.1(0.3) & h \\
W Crt	& 0.4120 & -0.54 & 0.05  &  59 $\pm$ 1 & 0.35-0.8 &1170 $\pm$ 25(90)  & 3.7 $\pm$ 0.1(0.4) & c,d \\
UU Vir  & 0.4756 & -0.87 & 0.03  &  -8 $\pm$ 1 & 0.1-0.9  & 880 $\pm$ 20(40)  & 4.6 $\pm$ 0.2(0.4) & g \\
BB Pup	& 0.4805 & -0.64 & 0.10  & 130 $\pm$ 1 & 0.0-1.0  &1520 $\pm$ 70(30)  & 4.3 $\pm$ 0.2(0.1) & c,d \\
RR Lyr	& 0.5668 & -1.39 & 0.07  & -72 $\pm$ 1 & 0.0-1.0  & 270 $\pm$ 25(25)  & 5.2 $\pm$ 0.5(0.5) & i,f \\
RV Oct	& 0.5711 & -1.71 & 0.13  & 136 $\pm$ 1 & 0.0-0.85 & 960 $\pm$ 20(40)  & 5.3 $\pm$ 0.2(0.2) & c,d \\
WY Ant	& 0.5743 & -1.48 & 0.05  & 204 $\pm$ 1 & 0.2-0.9  &1120 $\pm$ 35(60)  & 5.7 $\pm$ 0.2(0.4) & c,d \\
SS Leo	& 0.6263 & -1.50 & 0.01  & 161 $\pm$ 1 & 0.0-1.0  &1620 $\pm$ 40(70)  & 7.2 $\pm$ 0.2(0.4) & h \\
\enddata	
\tablenotetext{a}{Mean radial velocity (kms$^{-1}$). 
\hspace*{2.5mm} $^b$ First overtone pulsator.  
References:
(c) Skillen et al. (1993a);
(d) Skillen et al. (1993b);
(e) Jones, Carney \& Latham (1988a);
(f) Manduca \& Bell (1981);
(g) Jones, Carney \& Latham (1988b);
(h) Fernley et al. (1990);
(i) Wilson (1953).
}	
\end{deluxetable}

\clearpage
\begin{figure}
\plotone{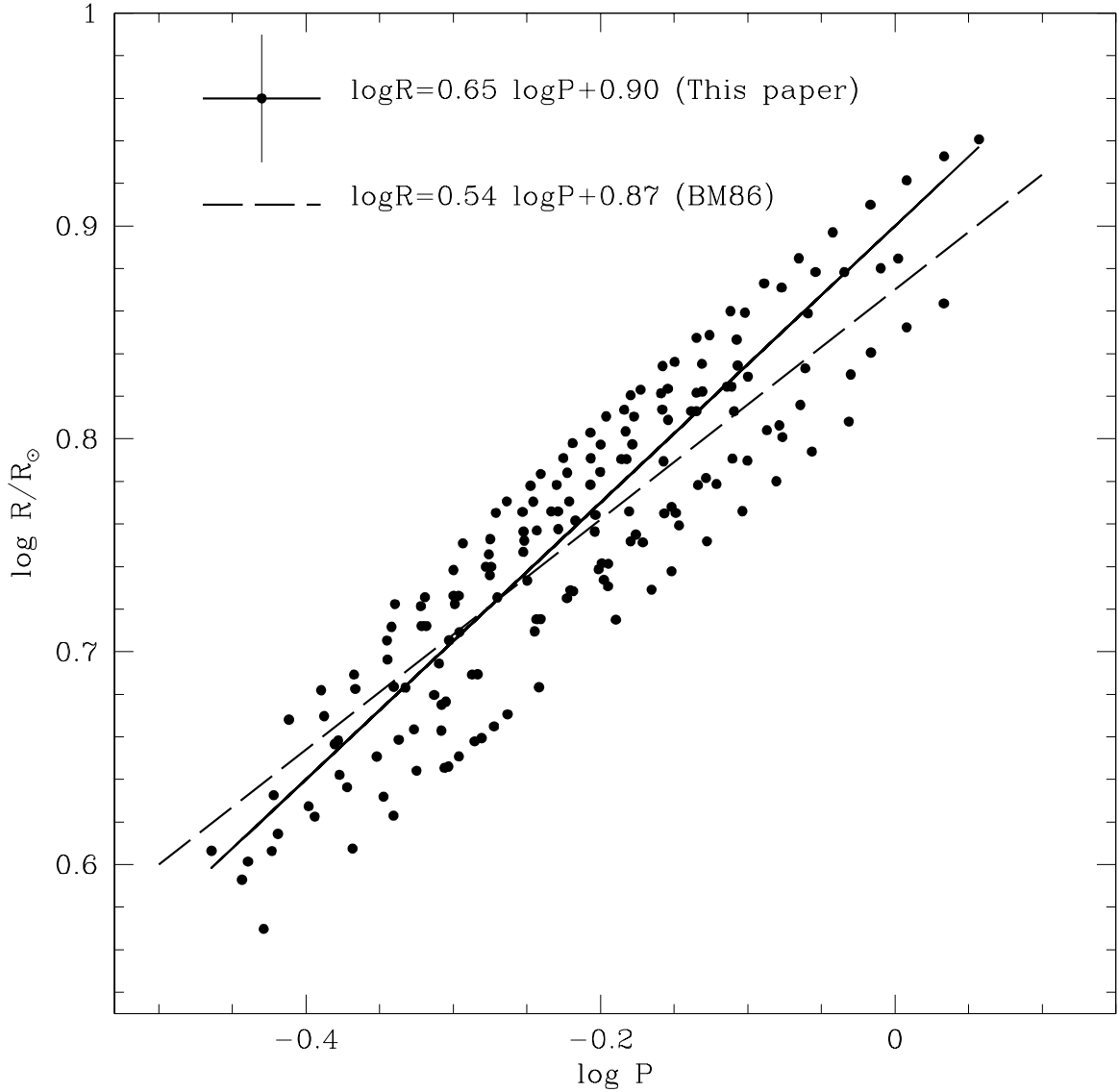}
\caption{Predicted radii for fundamental RR Lyrae models as a function of the 
logarithmic period. Solid line shows the predicted linear regression over the 
entire set of models, while the dashed one the empirical PR relation for 
Type II Cepheids derived by Burki \& Meylan (1986).      
The vertical error bar plotted in the left top corner shows the large
intrinsic dispersion of predicted radii.\label{fig1}}
\end{figure}

\begin{figure}
\plotone{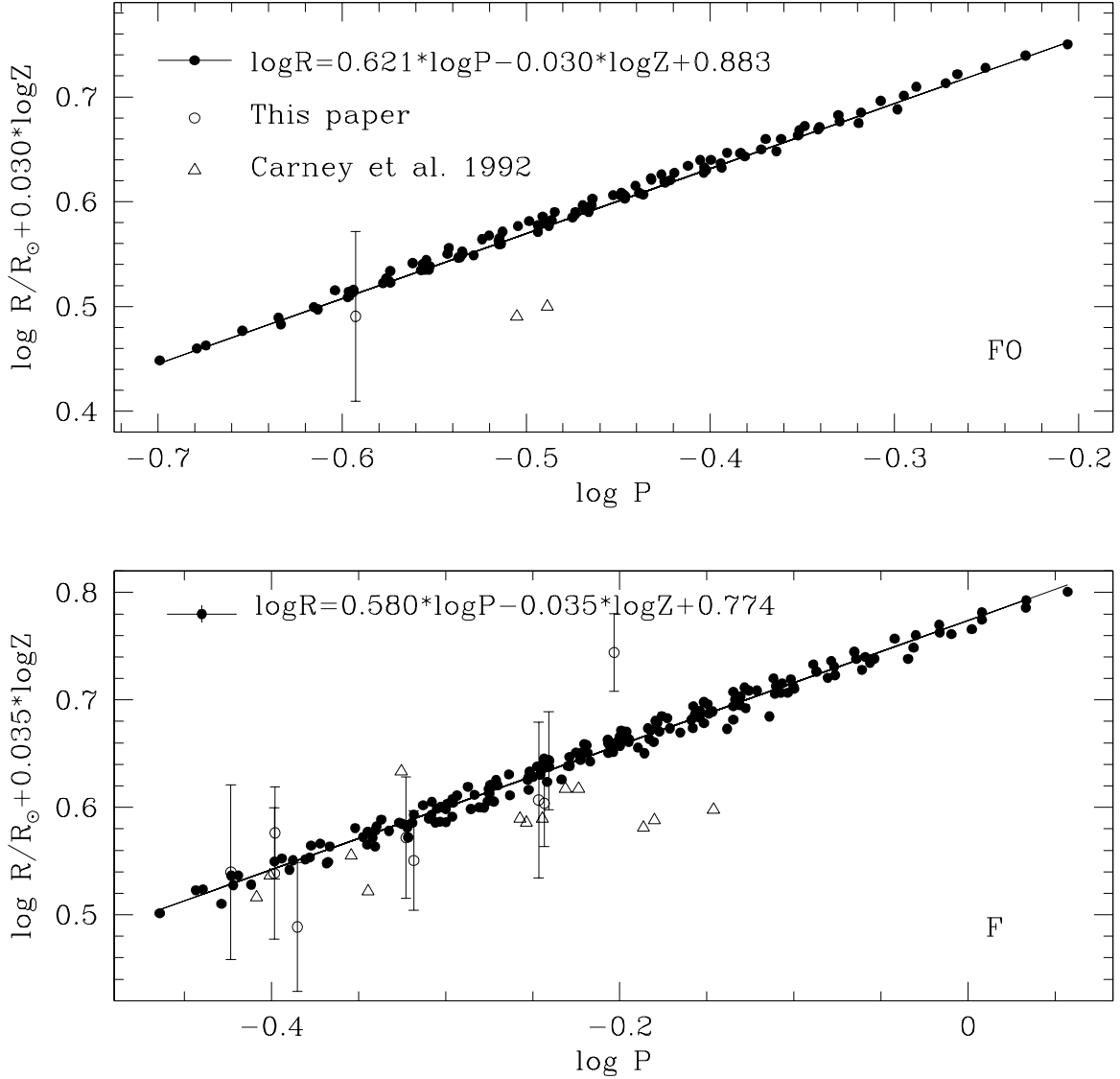}
\caption{Period-Radius-Metallicity relation for first overtone (top) and 
fundamental (bottom) RR Lyrae models projected onto a two-dimensional plane. 
Filled and open circles display predicted and new radius estimates, while 
triangles the radius determinations by Carney et al. (1992). Empirical radii 
have been plotted assuming metal abundances by Fernley et al. (1998) and 
$Z_\odot=0.02$. Individual error bars account for both random and systematic 
errors. \label{fig2}}
\end{figure}

\begin{figure}
\plotone{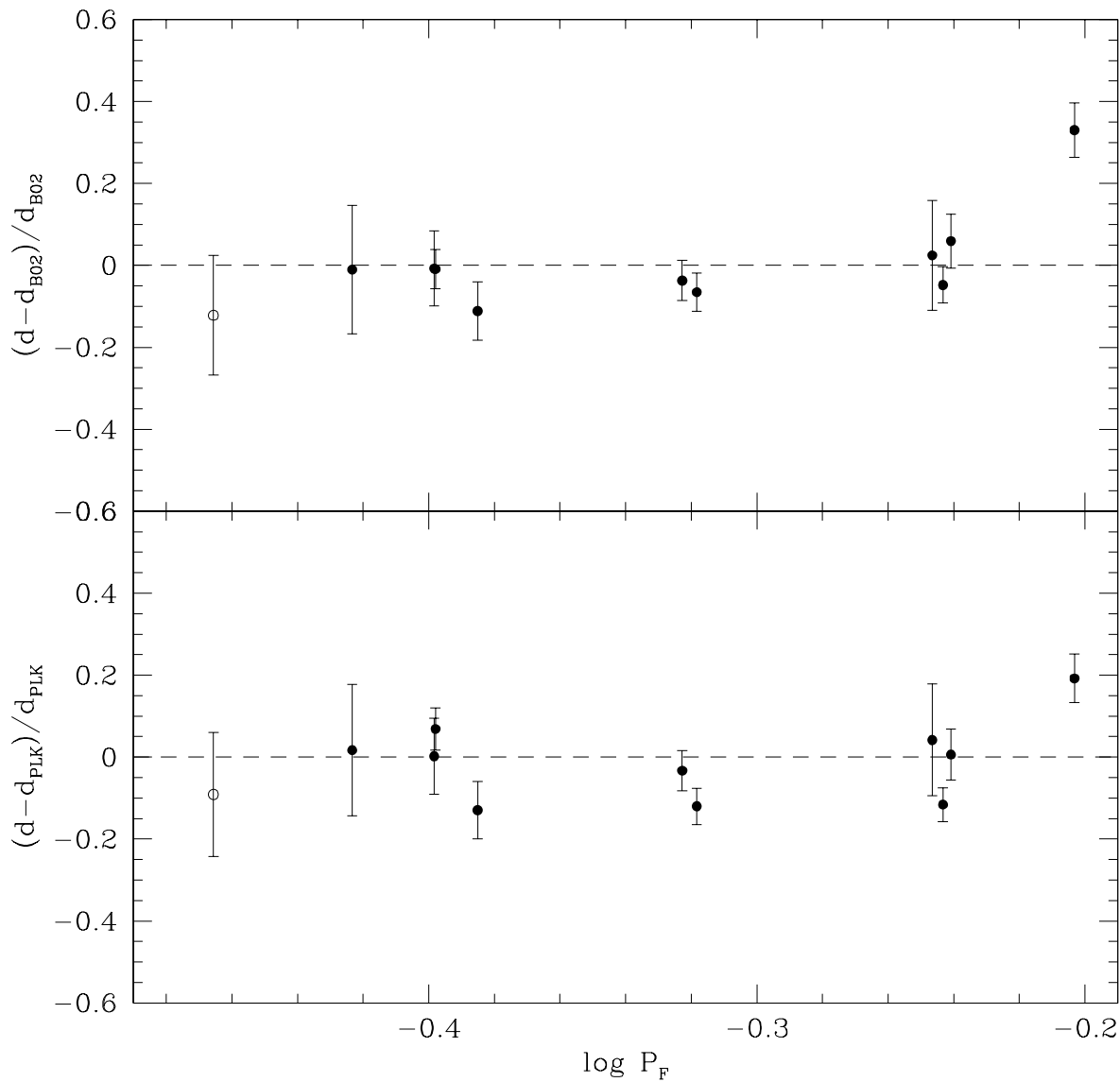}
\caption{Top panel - Relative difference between current distances and distances 
estimated using the calibration of the $M_V vs [Fe/H]$ relation provided by 
B02. Individual error bars account for both random and 
systematic errors. The period of DH Peg (open circle) was fundamentalized. 
Bottom panel - Same as the top, but with distances estimated using the 
$PLZ_{\rm K}$ relation provided by B03.\label{fig3}}
\end{figure}

\end{document}